\documentclass[11pt]{article}\textwidth 6.5in\textheight 9in
\usepackage{amssymb}\usepackage[colorlinks]{hyperref}\usepackage{color}
	\usepackage{stmaryrd}\usepackage{mathrsfs}
		\usepackage{graphicx}
	\usepackage{amsmath}
\usepackage{caption}
    \usepackage{subcaption}
    \usepackage{listings}
 	\usepackage{amsthm}

\begin{document}
\setlength{\captionmargin}{27pt}
\newcommand\hreff[1]{\href {http://#1} {\small http://#1}}
\newcommand\trm[1]{{\bf\em #1}} \newcommand\emm[1]{{\ensuremath{#1}}}

\newcommand\prf{\paragraph{Proof.}}\renewcommand\qed{\hfill\emm\blacksquare}

\newtheorem{thr}{Theorem} 
\newtheorem{lmm}{Lemma}
\newtheorem{cor}{Corollary}
\newtheorem{con}{Conjecture} 
\newtheorem{prp}{Proposition}

\theoremstyle{definition}
\newtheorem{blk}{Block}
\newtheorem{dff}{Definition}
\newtheorem{asm}{Assumption}
\newtheorem{rmk}{Remark}
\newtheorem{clm}{Claim}
\newtheorem{exm}{Example}

\newcommand\Ks{\mathbf{Ks}} 
\newcommand{\ab}{a\!b}
\newcommand{\yx}{y\!x}
\newcommand{\yux}{y\!\underline{x}}

\newcommand\floor[1]{{\lfloor#1\rfloor}}\newcommand\ceil[1]{{\lceil#1\rceil}}

\newcommand{\lea}{<^+}
\newcommand{\gea}{>^+}
\newcommand{\eqa}{=^+}

\newcommand{\lel}{<^{\log}}
\newcommand{\gel}{>^{\log}}
\newcommand{\eql}{=^{\log}}

\newcommand{\lem}{\stackrel{\ast}{<}}
\newcommand{\gem}{\stackrel{\ast}{>}}
\newcommand{\eqm}{\stackrel{\ast}{=}}

\newcommand\edf{{\,\stackrel{\mbox{\tiny def}}=\,}}
\newcommand\edl{{\,\stackrel{\mbox{\tiny def}}\leq\,}}
\newcommand\then{\Rightarrow}

\newcommand\C{\mathbf{C}} 

\renewcommand\chi{\mathcal{H}}
\newcommand\km{{\mathbf {km}}}\renewcommand\t{{\mathbf {t}}}
\newcommand\KM{{\mathbf {KM}}}\newcommand\m{{\mathbf {m}}}
\newcommand\md{{\mathbf {m}_{\mathbf{d}}}}\newcommand\mT{{\mathbf {m}_{\mathbf{T}}}}
\newcommand\K{{\mathbf K}} \newcommand\I{{\mathbf I}}

\newcommand\II{\hat{\mathbf I}}
\newcommand\Kd{{\mathbf{Kd}}} \newcommand\KT{{\mathbf{KT}}} 
\renewcommand\d{{\mathbf d}} 
\newcommand\D{{\mathbf D}}

\newcommand\w{{\mathbf w}}
\newcommand\Cs{\mathbf{Cs}} \newcommand\q{{\mathbf q}}
\newcommand\E{{\mathbf E}} \newcommand\St{{\mathbf S}}
\newcommand\M{{\mathbf M}}\newcommand\Q{{\mathbf Q}}
\newcommand\ch{{\mathcal H}} \renewcommand\l{\tau}
\newcommand\tb{{\mathbf t}} \renewcommand\L{{\mathbf L}}
\newcommand\bb{{\mathbf {bb}}}\newcommand\Km{{\mathbf {Km}}}
\renewcommand\q{{\mathbf q}}\newcommand\J{{\mathbf J}}
\newcommand\z{\mathbf{z}}

\newcommand\B{\mathbf{bb}}\newcommand\f{\mathbf{f}}
\newcommand\hd{\mathbf{0'}} \newcommand\T{{\mathbf T}}
\newcommand\R{\mathbb{R}}\renewcommand\Q{\mathbb{Q}}
\newcommand\N{\mathbb{N}}\newcommand\BT{\{0,1\}}
\newcommand\FS{\BT^*}\newcommand\IS{\BT^\infty}
\newcommand\FIS{\BT^{*\infty}}
\renewcommand\S{\mathcal{C}}\newcommand\ST{\mathcal{S}}
\newcommand\UM{\nu_0}\newcommand\EN{\mathcal{W}}
\renewcommand\H{\mathbf{H}}

\newcommand{\supp}{\mathrm{Supp}}

\newcommand\lenum{\lbrack\!\lbrack}
\newcommand\renum{\rbrack\!\rbrack}

\newcommand\h{\mathbf{h}}
\renewcommand\qed{\hfill\emm\square}
\renewcommand\i{\mathbf{i}}
\newcommand\p{\mathbf{p}}
\renewcommand\q{\mathbf{q}}
\title{\vspace*{-3pc} Uniform Tests and  Algorithmic Thermodynamic Entropy}

\author {Samuel Epstein\footnote{JP Theory Group. samepst@jptheorygroup.org}}

\maketitle
\begin{abstract}
We prove that given a computable metric space and two computable measures, the set of points that have high universal uniform test scores with respect to the first measure will have a lower bound with respect to the second measure.
This result is transferred to thermodynamics, showing that algorithmic thermodynamic entropy must oscillate in the presence of dynamics. Another application is that outliers will become emergent in computable dynamics of computable metric spaces.
\end{abstract}
\section{Introduction}
The study of randomness of computable metric spaces can be seen in the works of \cite{HoyrupRo09,Gacs21}. These spaces are important because physical random phenomena are modeled using infinite objects, and not the Cantor space. In this paper, we prove a result in computable metric spaces and then transfer it to physics, in particular, thermodynamics. For definitions in this introduction, we use \cite{HoyrupRo09}. A computable metric space $\mathcal{X}$ is a metric space with a dense set of ideal points on which the distance function is computable. A computable probability is defined by a computable sequence of converging points in the corresponding space of Borel probability measures, $\mathcal{M}(\mathcal{X})$, over $\mathcal{X}$. A uniform test $T$ takes in a description of a probability measure $\mu$ and produces a lower computable $\mu$ test, with $\int_\mathcal{X}T^\mu d\mu\leq 1$.
There exists a universal test, $\t$, where for any uniform test $T$ there is a $c_T\in\N$ where $c_T\t>T$. We extend Theorem 3 of \cite{EpsteinDerandom22} to computable metric spaces.\\

\noindent\textbf{Theorem.} \textit{
Given computable probability measures $\mu$ and $\lambda$, non-atomic $\lambda$, over a computable metric space $\mathcal{X}$ and universal uniform test $\t$, there is $c\in\N$ where for all $n$, $\lambda(\{\alpha:\t_\mu(\alpha)>2^n\})>2^{-n-\K(n)-c}$.}

\subsection{Thermodynamics}

The above theorem can be applied to thermodynamics. Classical thermodynamics is the study of substances and changes to their properties such as volume, temperature, and pressure. Substances, such as a gas or a liquid, is modeled as a point in a phase space. The phase space, $\mathcal{X}$, is modeled by a computable metric space, \cite{HoyrupRo09}, and a volume measure $\mu$, is modeled by a computable (not necessarily probabilistic) positive measure over $\mathcal{X}$. The dynamics are modeled by a one dimensional transformation group $G^t$, indexed by $t\in\R$. Due to Louville's theorem, the dynamics are measure-preserving, where $\mu(G^tA)=\mu(A)$, for all Borel sets $A\subseteq\mathcal{X}$. 

Whether quantum or classical, the known laws of physics are reversible. Thus the dynamics $G$ of our system are also reversible, in that if $\beta =G^t\alpha$, then there is some $t'$ such that the original state can found with $\alpha = G^{t'}\beta$. Thus if given a set of particles with position and velocity, by reversing the velocities, a previous state can be found. This is contradiction to the second law of thermodynamics, which states,
\begin{quote}
\textit{The total entropy of a system either increases or remains constant in any spontaneous process; it never decreases.}
\end{quote}
This conforms to our experiences of broken vases never reforming. To reconcile this difference, Boltzmann introduced \textit{macro-states}, $\Pi_i$, indexed by $i\in\N$, which groups states together by macroscopic parameters, with corresponding Boltzmann entropy $S(\Pi_i)=k_B\ln\mu(\Pi_i)$. By definition, a vast majority of typical states will  experience an increase in Boltzmann entropy.

In \cite{Gacs94}, coarse grained entropy was introduced as an algorithmic update to Boltzmann entropy. This formulation was made to be independent of the choice of parameters of the macro state. In this paper, we introduce a modified version of coarse grained entropy. We also model the thermodynamic entropy of a micro-state with algorithmic methods. The micro-state of a system  contains the information of the entire physical  state. For example, the microstate of a system of $N$ molecules is a point
$$(q_1,\dots,q_{3N},p_1,\dots,p_{3N})\in\R^{6N}$$
where $q_i$ are the position coordinates and $p_i$ are the momentum coordinates. The set of states, $\R^{3N}$ is a computable metric space. To model the entropy of the state, we use slight variant to algorithmic fine-grained entropy $H_\mu$ in \cite{Gacs94}, using symbol $\H_\mu$. This entropy measure captures the level of disorder of the state. Continuing the example above, if all the particles are at rest, then the thermodynamic entropy of the state of 
$$
(q_1,\dots,q_{3N},0,\dots,0)
$$
is expected to be very low.
\begin{quote}
\textit{The evolution of the system will be thermodynamic like if it spends most of the time close to its maximum value, from
which it exhibits frequent small fluctuations and rarer large fluctuations.}
\end{quote}
In this paper, using the algorithmic definition of thermodynamic entropy, $\H_\mu$, we prove that such fluctuations \textit{have to} occur, and the greater the fluctuation, the lesser its measure. The symbol $\mu$ is a Borel measure representing volume of the phase space. In this paper, we show that thermodynamic entropy $\H_\mu$ must oscillate in the presence of dynamics.\\

\noindent\textbf{Theorem.} \textit{Let $L$ be the Lebesgue measure over $\R$ and $\alpha\in\mathcal{X}$, with finite mutual information with the halting sequence. For transformation group $G^t$ acting on computable metric space $\mathcal{X}$, there are constants $c_1$ and $c_2$ with $2^{-n-\K(n)-c_1}<L\{t\in [0,1]:\H_\mu(G^t\alpha) < \H_\mu(\alpha)-n\}< 2^{-n+c_2}$.}

\subsection{Outliers in Dynamics}

An application of the thermodynamics theorem is that when outlier scores are modeled using $\t_\mu$, then they will become emergent in dynamics of transformation groups $G^t$. This parallels the case of Cantor, as seen in \cite{EpsteinDerandom22,EpsteinDynamics22}.\\

\noindent \textbf{Corollary.} \textit{Let $L$ be the Lebesgue measure over $\R$, $(\mathcal{X},\mu)$ be a computable probability space, $\alpha\in\mathcal{X}$, with finite mutual information with the halting sequence. For transformation group $G^t$ acting on $\mathcal{X}$, there is a constant $c$ with $L\{t\in [0,1]:\t_\mu(G^t\alpha) > 2^n\}> 2^{-n-\K(n)-c}$.}

\section{Computable Probability Spaces}
\label{sec:probspace}
The first main result of this paper uses computable metric spaces and computable probability measures from \cite{HoyrupRo09}. Some constructs need changes, which we present in later sections. But in this section we show the definitions, lemmas, and theorems that are directly taken from \cite{HoyrupRo09}. If a theorem or lemma is presented without a proof, then it can be found in \cite{HoyrupRo09}
\begin{dff}
A computable metric space consists of a triple $(\mathcal{X},\mathcal{S},d)$, where
\begin{itemize}
\item $\mathcal{X}$ is a separable complete metric space.
\item $\mathcal{S}$ is an enumerable list of dense ideal points $\mathcal{S}$ in $\mathcal{X}$.
\item $d$ is a distance metric that is uniformly computable over points in $\mathcal{S}$.
\end{itemize}
\end{dff}
For $x\in\mathcal{X}$, $r\in\Q_{>0}$ a ball is $B(x,r)=\{y:d(x,y)<r\}$. The ideal points induce a sequence of enumerable ideal balls $B_i = \{ B(s_i,r_j) : s_i\in\mathcal{S}, r_j\in \Q_{>0}\}$. A sequence of ideal points $\{x_n\}\subseteq X$ is said to be a fast Cauchy sequence if $d(x_n,x_{n+1})<2^{-n}$ for all $n\in\N$. A point $x$ is computable there is a computable fast Cauchy sequence converging to $x$. Each computable function $f$ between computable metric spaces $\mathcal{X}$ and $\mathcal{Y}$ has an algorithm $\mathfrak{A}$ such that if $f(x)=y$ then for all fast Cauchy  sequences $\overrightarrow{x}$ for $x$, $\mathfrak{A}(\overrightarrow{x})$ outputs an encoding of a fast Cauchy sequence for $y$.

\begin{dff}
Lower computable functions $f\in \mathcal{F}$ have algorithms that enumerate $\{(B_i,r_i)\}$, where $B_i$ is an ideal ball and $r_i\in\Q_{> 0}$, and $f(x) = \sup\{r_i:x\in B_i\}$.
\end{dff}
The computable metric space of all Borel probability measures over $\mathcal{X}$ is $\mathcal{M}(\mathcal{X})$. If $\mathcal{X}$ is separable and compact then so is $\mathcal{M}(\mathcal{X})$. The ideal points of $\mathcal{M}(\mathcal{X})$ are $\mathcal{D}$, the set of probability measures that are concentrated on finitely many points with rational values. The distance metric on $\mathcal{M}(\mathcal{X})$ is the \textit{Prokhorov metric}, defined as follows.
\begin{dff}[Prokhorov metric]
\label{dff:Prokhorov}
$$
\pi(\mu,\nu)=\inf\left\{\epsilon\in\R^+:\mu(A)\leq \nu(A^\epsilon)\textrm{ for Borel set }A\right\},
$$ 
where $A^\epsilon=\{x:d(x,A)<\epsilon\}$.
\end{dff}
\begin{thr}
\label{thr:compequivmeasure}
Given a probability measure $\mu\in\mathcal{M}(\mathcal{X})$, the following are equivalent.
\begin{enumerate}
\item $\mu$ is computable.
\item $\mu(B_{i_1}\cup\dots\cup B_{i_k})$ is lower semi-computable uniformly in $\langle i_1,\dots,i_k\rangle$.
\item $\int d\mu:\mathcal{F}\rightarrow \R_{\geq 0}$ is lower semi-computable.
\end{enumerate}
\end{thr}
\begin{dff}$ $\\
\label{dff:binary}
\vspace*{-0.5cm}
\begin{enumerate}
\item A \textit{constructive $G_\delta$-set} is a set of the form $\bigcap_nU_n$ where $\{U_n\}$ is a sequence of uniformly r.e. open sets.
\item A \textit{computable probability space} is a pair $(\mathcal{X},\mu)$, where $\mathcal{X}$ is a computable metric space and $\mu$ is a Borel probability measure on $\mathcal{X}$.
\item Let $(\mathcal{X},\mu)$ be a computable probability space and $\mathcal{Y}$ a computable metric space. A function $f:D_f\subset(\mathcal{X},\mu)\rightarrow\mathcal{Y}$ is \textit{almost computable} if it is computable on a constructive $G_\delta$-set ($D_f$) of $\mu$-measure one.
\item A \textit{morphism} of computable probability spaces $Q:(\mathcal{X},\mu)\rightarrow (\mathcal{Y},\nu)$ is an almost computable measure-preserving function $Q:D_Q\subset\mathcal{X}\rightarrow\mathcal{Y}$, where $\mu(Q^{-1}(A))=\nu(A)$ for all Borel sets $A$. An \textit{isomorphism} $(Q,R)$ is a pair of morphisms such that $Q\circ R = \mathrm{id}$ on $R^{-1}(D_Q)$ and $R\circ Q=\mathrm{id}$ on $Q^{-1}(D_R)$. 
\item  A \textit{binary representation} of a computable probability space $(\mathcal{X},\mu)$ is a pair $(\delta,\mu_\delta)$ where $\mu_\delta$ is a computable probability measure on $\IS$ and $\delta:(\IS,\mu_\delta)\rightarrow (\mathcal{X},\mu)$ is a surjective morphism such that, calling $\delta^{-1}(x)$ the set of expansions of $x\in X$:
\begin{itemize}
\item There is a dense full-measure constructive $G_\delta$-set $D$ of points having a unique expansion.
\item $\delta^{-1}:D\rightarrow \delta^{-1}(D)$ is computable.
\item $(\delta,\delta^{-1})$ is an isomorphism.
\end{itemize}
\end{enumerate}
\end{dff}
\begin{thr}
\label{thr:binary}
Every computable probability space $(\mathcal{X},\mu)$ has a binary representation.
\end{thr}
\begin{dff}
Given a probability measure $\mu\in\mathcal{M}(\mathcal{X})$, a $\mu$-randomness test is a $\mu$-constructive function $T\in\mathcal{F}$, such that $\int Td\mu\leq 1$. A uniform randomness test is a constructive function $T$ from $\mathcal{M}(\mathcal{X})$ to $\mathcal{F}$ such that $\int T^\mu d\mu\leq 1$.
\end{dff}
\begin{thr}$ $\\
\label{thr:equivuniform}
\vspace*{-0.5cm}
\begin{enumerate}
\item Let $\mu$ be a probability measure. For every $\mu$-randomness test $t$, there is a uniform randomness test $T:\mathcal{M}(\mathcal{X})\rightarrow\mathcal{F}$ with $T(\mu)=.5t$.
\item There is a universal uniform randomness test, that is a uniform test $\mathbf{t}$ such that for every uniform test $T$, there is a constant $c>0$ with $\t>cT$.
\end{enumerate}
\end{thr}

\section{Dual Binary Representation}
This paper introduces a new concept that is needed in the first theorem: dual binary representation. While a binary representation is a mapping from one computable probability space to the Cantor space, a dual binary representation maps two computable probability spaces to Cantor spaces, each sharing the same mapping.

\begin{dff}A set $A$ is \textit{almost decidable} with respect to probability measures $(\mu,\nu)$ if there are two. r.e. open sets $U$ and $V$ such that $U\subset A$, $V\subseteq A^\mathscr{C}$, $U\cup V$ is dense and has full $\mu$ and $\nu$ measure. We say the elements of a sequence $\{A_i\}$ are uniformly almost decidable with respect to $(\mu,\nu)$ if there are two sequences $\{U_i\}$ and $\{V_i\}$ of uniformly r.e. sets satisfying the above conditions.
\end{dff}
\begin{thr}
\label{thr:Baire}
On a computable metric space, every dense constructive $G_\delta$-set has a dense sequence of uniformly computable points.
\end{thr}
\begin{lmm}
\label{lmm:existalmost}
There is a sequence of $\{r_n\}$ of uniformly computable reals such that $\{B(s_i,r_n)\}_{i,n}$ is a basis of uniformly almost computable decidable balls, relative to $(\mu,\nu)$.
\end{lmm}
\begin{prf}
Define $U_{\langle i,k\rangle} = \{r\in\R_{>0}:\mu(\overline{B}(s_i,r))<\mu(B(s_i,r))+1/k\}$. By computability of $\mu$, this is a r.e. open subset of $\R_{>0}$ uniformly in $\langle i,k\rangle$. Let $W_{\langle i,k\rangle} = \{r\in\R_{>0}:\nu(\overline{B}(s_i,r))<\nu(B(s_i,r))+1/k\}$, which is also an r.e. open subset of $\R_{> 0}$. They both are dense in $\R_{>0}$. The spheres $S_r = \overline{B}(s_i,r)\setminus B(s_i,r)$ are disjoint for different radii and $\mu$ and $\nu$ are finite, so the set of $r$ for which $\mu(S_r)\geq 1/k$ or $\nu(S_r)\geq 1/k$ is finite. Let $V_{\langle i,j\rangle} = \R_{>0}\setminus\{d(s_i,s_j)\}$ be a dense r.e. open set, uniformly in $\langle i,j\rangle$. Then by Theorem \ref{thr:Baire}, the dense constructive $G_\delta$-set
$$
\bigcap_{\langle i,k\rangle}U_{\langle i,k\rangle}\cap \bigcap_{\langle i,k\rangle}W_{\langle i,k\rangle}\cap
\bigcap_{\langle i,j\rangle}V_{\langle i,j\rangle}
$$
contains a sequence $\{r_n\}$ of uniformly computable reals numbers which is dense in $\R_{>0}$. For any $s_i$ and $r_n$, $B(s_i,r_n)$ is almost decidable, relative to $(\mu,\nu)$. Thus $\{(B_i,r_n)\}_{i,n}$ is a basis of uniformly almost computable decidable balls, relative to $(\mu,\nu)$.\qed
\end{prf}

\begin{dff}
A dual probability space $(\mathcal{X},\mu,\nu)$ is a computable metric space $\mathcal{X}$ and two computable Borel probability measures, $\mu$ and $\nu$, over $\mathcal{X}$.
\end{dff}
\begin{dff}
A dual binary representation of a dual probability space $(\mathcal{X},\mu,\nu)$ is a tuple $(\delta,\mu_\delta,\nu_\delta)$ where $\mu_\delta$ and $\nu_\delta$ are computable probability measures on $\IS$ and $\delta:(\IS,\mu_\delta)\rightarrow (\mathcal{X},\mu)$ and $\delta:(\IS,\nu_\delta)\rightarrow (\mathcal{X},\nu)$ are surjective morphisms. Denoting $\delta^{-1}(x)$ to be the set of expansion of $x\in X$: 
\begin{itemize}
\item There is a dense full-measure constructive $G_\delta$-set $D$ of points have a unique expansion.
\item $\delta^{-1}:D\rightarrow \delta^{-1}(D)$ is computable.
\item $(\delta,\delta^{-1})$ is an isomorphism.
\end{itemize}
\end{dff}

\begin{thr}
\label{thr:dual}
Every dual probability space $(\mathcal{X},\mu,\nu)$ has a dual binary representation.
\end{thr}
\begin{prf}
This proof follows identically to the proof of Theorem 5.1.1 in \cite{HoyrupRo09}, using Lemma \ref{lmm:existalmost} instead of Lemma 5.1.1, and noting that $\mu_\delta = \mu\circ b^{-1}$ and $\nu_\delta = \nu\circ b^{-1}$, where $b$ is defined in the proof.\qed
\end{prf}
\section{Universal Uniform Tests}

This section contains the first main result of the paper, Theorem \ref{thr:main1}. This section also includes further results needed for the proof of this theorem. Lemma \ref{lmm:morphism}, is derived from Proposition 6.2.1 from \cite{HoyrupRo09}. We also include definitions and theorems about the Cantor space, including Theorem \ref{thr:InfExt}.

\begin{lmm}
\label{lmm:morphism}
Let $Q:D\subset \mathcal{X}\rightarrow \mathcal{Y}$ be a morphism of computable probability spaces $(\mathcal{X},\mu)$ and $ (\mathcal{Y},\nu)$, with universal tests $\t_\mu$ and $\t_\nu$. There is a $c\in \N$ with the following properties. If $x\in \mathcal{X}$ and $\t_\mu(x)<\infty$, then $Q(x)$ is defined and $\t_\nu(Q(x))<c\t_\mu(x)$.
\end{lmm}
\begin{prf}
The proof is a slight modification to Proposition 6.2.1 in \cite{HoyrupRo09}. So, assuming $\t_\mu(x)<\infty$, then $x$ is a random point then $x\in D$, because due to Lemma 6.2.1 in \cite{HoyrupRo09}, every random point lies in every r.e. open set of full measure, and $D$ is an intersection of full-measure r.e open sets. Thus $Q(x)$ is defined.
 
 Let $\mathscr{A}$ be any algorithm lower semi-computing the function $\t_\nu\circ Q:D\rightarrow \R^\infty_{\geq 0}$. This algorithm can be converted into a lower computable function $f_\mathscr{A}:\mathcal{X}\rightarrow \R^\infty_{\geq 0}$ by feeding all finite prefixes of fast Cauchy sequences to $Q$ and enumerating all resultant outputted ideal balls and seeing which outputted ideal balls are in the ideal balls of those enumerated by $\t_\nu$. Since $\mu(D)=1$, $\int\t_\nu\circ Qd\mu$ equals $\int f_{\mathscr{A}}d\mu$. As $Q$ is measure-preserving, $\int \t_\nu\circ Qdu=\int \t_\nu d\nu\leq 1$. Hence $f_\mathscr{A}$ is a $\mu$-test, with $f_\mathscr{A}<c\t_\mu$ for some $c\in \N$. Thus $\t_\nu(Q(x))=f_\mathscr{A}(x) < c\t_\mu(x)$.\qed
\end{prf}
\begin{cor}
Let $(Q,R):(\mathcal{X},\mu)\rightleftarrows (\mathcal{Y},\nu)$ be an isomorphism of computable probability spaces, with universal tests $\t_\mu$ and $\t_\nu$. Then there is a $c\in \N$ where $\t_\nu(Q(x))= \t_\mu(x)\pm c$ and $\t_\mu(R(y))=\t_\nu(y)\pm c$.
\end{cor}

The algorithmic probability is $\m(x|y)$. The conditional prefix free complexity is $\K(x|y)$. Symmetric mutual information is $\I(x:y)=\K(x)+\K(y)-\K(x,y)$. The halting sequence is $\ch\in\IS$. For positive real functions $f$, by ${\lea}f$, ${\gea}f$, ${\eqa}f$, denote ${\leq}\,f{+}O(1)$, ${\geq}\,f{-}O(1)$, ${=}\,f{\pm}O(1)$, respectively. In addition ${\lem}f$, ${\gem}f$, and ${\eqm}f$ denote $<f/O(1)$, $>f/O(1)$ and $=f*/O(1)$, respectively. 
\begin{dff}
\label{dff:defrand}
For computable probability $P$, $\langle P\rangle\in\FS$ is the smallest program that can compute $P(x\IS)$, uniformly computable $x\in\FS$.  The deficiency of randomness of an infinite sequence $\alpha\in\IS$ with respect to a computable probability measure $P$ over $\IS$ is defined to be $$\D(\alpha|P,x)=\log\sup_n\m(\alpha[0..n]|\langle P\rangle,x)/P(\alpha[0..n]).$$ 
We have $\D(\alpha|P)=\D(\alpha|P,\emptyset)$. By \cite{Gacs21}, $2^\D$ is a lower-computable $P$-test, in that $\int_{\IS} 2^{\D(\alpha|P)}dP(\alpha)=O(1)$, lower computed by program of size $\K(\langle P\rangle)$. Thus since $\t_P$ is a universal uniform test, $\t_P(\alpha)\gem \m(\langle P\rangle)2^{\D(\alpha|P)}$.
\end{dff}
\begin{dff}
The information term between infinite sequences is\\ $\I(\alpha:\beta)=\log\sum_{x,y\in\FS}\m(x|\alpha)\m(y|\beta)2^{\I(x:y)}$ \cite{Levin74}. $\I(f(\alpha):\beta)\lea \I(\alpha:\beta)+\K(f)$.
\end{dff}

\begin{thr}[\cite{Vereshchagin21,Levin74,Geiger12}]
	\label{thr:coninfo}
	$\Pr_{\mu}(\I(\alpha:\ch)>n)\lem 2^{-n+\K(\mu)}$, where $\K(\mu)$ is the size of the smallest program that can compute $\mu(x\IS)$, uniformly in $x\in\FS$.
\end{thr}

\begin{thr}[\cite{Epstein21}]
	\label{thr:InfExt} 
	For computable probability measure $P$ over $\IS$, for $Z\subseteq \IS$,  if $\N\ni s< \log\sum_{\alpha\in Z}2^{\D(\alpha|P)}$, then $s<\sup_{\alpha\in Z}\D(\alpha|P)\,{+}\,\I(\langle Z\rangle:\ch)+O(\K(s)+\log\I(\langle Z\rangle:\ch)+\K(P))$.
\end{thr}
\begin{thr}[\cite{EpsteinDerandom22}]
\label{thr:revisited}
For computable measures $\mu$ and non-atomic $\lambda$ over $\IS$ and $n\in\N$,\\ $\lambda\{\alpha :\D(\alpha|\mu)>n\}> 2^{-n-\K(n,\mu,\lambda)-O(1)}$. 
\end{thr}

\begin{thr}
\label{thr:main1}
Given computable probability measures $\mu$ and $\lambda$, non-atomic $\lambda$, over a computable metric space $\mathcal{X}$ and universal uniform test $\t$, there is $c\in\N$ where for all $n$, $\lambda(\{\alpha:\t_\mu(\alpha)>2^n\})>2^{-n-\K(n)-c}$.
\end{thr}
\begin{prf}
By Theorem \ref{thr:binary}, fix a dual binary representation $(\delta,\lambda_\delta,\mu_\delta)$ for dual probability space $(\mathcal{X},\lambda,\mu)$. Note that $\delta$ is a measure-preserving transform, where $\lambda(A) = \lambda_\delta(\delta^{-1}(A))$ for all Borel sets $A$. Due to Lemma \ref{lmm:morphism}, there is some $c'\in\N$ where 
\begin{align}
\nonumber
=&\lambda(\{\beta:\t_\mu(\beta)\leq 2^n\})\\
\nonumber
=&\lambda_\delta(\delta^{-1}(\{\beta:\t_\mu(\beta)\leq 2^n\}))\\
\nonumber
\leq&\lambda_\delta(\delta^{-1}(\{\beta:\t_{\mu_\delta}(\delta^{-1}(\beta))\leq 2^{n+c'}\}))\\
\nonumber
=&\lambda_\delta(\delta^{-1}(\{\beta\in\delta(\{\alpha:\t_{\mu_\delta}(\alpha)\leq 2^{n+c'}\})\})\\
\nonumber
=&\lambda_\delta(\delta^{-1}(\delta(\{\alpha:\t_{\mu_\delta}(\alpha)\leq 2^{n+c'}\}))\\
\label{eq:deltalambda}
\leq&\lambda_\delta(\{\alpha:\t_{\mu_\delta}(\alpha)\leq 2^{n+c'}\}).
\end{align}
Combining Equation and \ref{eq:deltalambda}, and Definition \ref{dff:defrand}, we get, (by updating $c'$)
\begin{align}
\nonumber
\lambda(\{\beta:\t_\mu(\beta)\leq 2^n\})\leq &\lambda_\delta\{\alpha:\D(\alpha|\mu_\delta)\leq n+c'\}.
\end{align}
By Theorem \ref{thr:revisited}, we get for a constant $c''$ dependent on $\lambda$ and $\mu$,
\begin{align*}
\lambda(\{\beta:\t_\mu(\beta)\leq 2^n\})< &1-2^{-n-\K(n)-c''}.\\
\end{align*}\qed
\end{prf}

\section{Computable Measure Theory}
In thermodynamics, the measure function representing the volume is not necessarily a probability measure. Thus the 
 results of Section \ref{sec:probspace} needs to be extended to nonnegative measures of arbitrary size to prove a result about thermodynamics. Let $(\R_{\geq 0},\Q_{\geq 0},d_\R)$ be the computable metric space where $\R_{\geq 0}$ is the complete separable metric space and nonnegative rationals $\Q_{\geq 0}$ consists of the ideal points. The distance function is $d_\R(x,y)=|x-y|$, which is obviously computable over $\Q_{\geq 0}$. The space of nonnegative Borel measures over a computable metric space is the space $\mathfrak{M}(\mathcal{X}) = \mathcal{M}(X)\times \R_{\geq 0}$, the product space of the space of probability measures of $\mathcal{X}$, $\mathcal{M}(\mathcal{X})$, with the space of nonnegative reals. The distance function of $\mathfrak{M}$ is $d_\mathfrak{M}((\mu,m),(\nu,n)) = \max\{\pi(\mu,\nu),d_\R(n,m)\}$, where $\pi$ is the Prokhorov metric (see Definition \ref{dff:Prokhorov}). The ideal points of $\mathfrak{M}(\mathcal{X})$ is the set of all finite points with nonnegative rational values. This definition is different from the ideal points in $\mathcal{M}(\mathcal{X})$ in that they don't have to sum to 1. The computable measures of $\mathfrak{M}(\mathcal{X})$ are its constructive points, with respect to a fast Cauchy description. From this definition, the results of Theorem \ref{thr:compequivmeasure} apply directly to arbitrary measures $\mu\in\mathfrak{M}(\mathcal{X})$.
\begin{cor}
\label{cor:compmeas}
Given an arbitrary measure $\mu\in\mathfrak{M}(\mathcal{X})$, the following are equivalent.
\begin{enumerate}
\item $\mu$ is computable.
\item $\mu(B_{i_1}\cup\dots\cup B_{i_k})$ is lower semi-computable uniformly in $\langle i_1,\dots,i_k\rangle$.
\item $\int d\mu:\mathcal{F}\rightarrow \R_{\geq 0}$ is lower semi-computable.
\end{enumerate}
\end{cor}
\begin{dff}$ $\\
\label{dff:measbinary}
\vspace*{-0.5cm}
\begin{enumerate}
\item A \textit{computable measure space} is a pair $(\mathcal{X},\mu)$, where $\mathcal{X}$ is a computable metric space, and $\mu$ is a Borel nonnegative measure on $\mathcal{X}$.
\item Let $(\mathcal{X},\mu)$ be a computable measure space and $\mathcal{Y}$ a computable metric space. A function $f:D_f\subset(\mathcal{X},\mu)\rightarrow\mathcal{Y}$ is \textit{almost computable} if it is computable on a constructive $G_\delta$-set  ($D_f$) of measure $\mu(\mathcal{X})$.
\item A \textit{morphism} of computable measure spaces $Q:(\mathcal{X},\mu)\rightarrow (\mathcal{Y},\nu)$ is an almost computable measure-preserving function $Q:D_Q\subset\mathcal{X}\rightarrow\mathcal{Y}$. An \textit{isomorphism} $(Q,R)$ is a pair of morphisms such that $Q\circ R = id$ on $R^{-1}(D_Q)$ and $R\circ Q=id$ on $Q^{-1}(D_R)$. 
\item  A \textit{binary representation} of a computable measure space $(\mathcal{X},\mu)$ is a pair $(\delta,\mu_\delta)$ where $\mu_\delta$ is a computable measure on $\IS$ and $\delta:(\IS,\mu_\delta)\rightarrow (\mathcal{X},\mu)$ is a surjective morphism such that, calling $\delta^{-1}(x)$ the set of expansions of $x\in X$:
\begin{itemize}
\item There is a dense constructive $G_\delta$-set $D$ of points having a unique expansion and $\mu(D)=\mu(\mathcal{X})$.
\item $\delta^{-1}:D\rightarrow \delta^{-1}(D)$ is computable.
\item $(\delta,\delta^{-1})$ is an isomorphism.
\end{itemize}
\end{enumerate}
\begin{cor}
Every computable measure space $(\mathcal{X},\mu)$ has a binary representation.
\end{cor}
This corollary has an identical proof to Theorem \ref{thr:binary}, which is from Theorem 5.1.1 in \cite{HoyrupRo09}.

\begin{dff}
Given an arbitrary measure $\mu\in\mathcal{M}(\mathcal{X})$, a $\mu$-randomness test is a $\mu$-constructive function $T\in\mathcal{F}$, such that $\int Td\mu\leq 1$. A uniform randomness test is a constructive function $T$ from $\mathcal{M}(\mathcal{X})$ to $\mathcal{F}$ such that $\int T^\mu d\mu\leq 1$.
\end{dff}
\begin{thr}$ $\\
\label{thr:equivuniform}
\vspace*{-0.5cm}
\begin{enumerate}
\item Let $\mu$ be a measure. For every $\mu$-randomnness test $t$, there is a uniform randomness test $T:\mathcal{M}(\mathcal{X})\rightarrow\mathcal{F}$ with $T(\mu)=.5t$.
\item There is a universal uniform randomness test, that is a uniform test $\mathbf{t}$ such that for every uniform test $T$, there is a constant $c>0$ with $\t>cT$.
\end{enumerate}
\end{thr}

\end{dff}
\section{Algorithmic Thermodynamic Entropy}
In this section, we prove the second main result of the paper, that thermodynamic entropy must oscillate in the presence of dynamics. Algorithmic thermodynamic entropy, $\H$, is formally defined as well as the transform group $G^t$ representing dynamics. Theorem \ref{thr:probcons} is a property about conservation of information with halting sequence. We also provide a new proof to a result in \cite{Gacs94} that states thermodynamics will decrease in only small measure.

\begin{dff}[Algorithmic Thermodynamic Entropy]
Given a computable metric space $\mathcal{X}$ and a nonnegative measure $\mu\in\mathfrak{M}(\mathcal{X})$ the algorithmic thermodynamic entropy is
	$\H_\mu(\alpha) = - \log\t_\mu(\alpha)$.
\end{dff}
\begin{dff}[Mutual Information with the Halting Sequence]
\label{dff:pointmuthalt}
An encoding of a fast Cauchy sequence $\overrightarrow{x}$ is $\langle \overrightarrow{x}\rangle\in\IS$, with $\langle \overrightarrow{x}\rangle=\langle x_1\rangle\langle x_2\rangle\dots$. Each point $x\in\mathcal{X}$ has a certain mutual information with the halting sequence $\I(x:\ch)=\inf \{\I(\langle \overrightarrow{x}\rangle:\ch):\langle \overrightarrow{x}\rangle\textrm{ is a fast Cauchy sequence for }x\}$.
\end{dff}

\begin{dff}[Computable Transformation Group]
\label{dff:transformgroup}
A one dimensional transformation group $G^t$, parameterized by $t\in\R$ over a measure space $(\mathcal{X},\mu)$ where each $G^t$ is a  homeomorphism of $\mathcal{X}$ onto itself, where $G^t(G^s(x)=G^{t+s}(x)$. And $G^tx$ is continuously simultaneously in $x$ and $t$. $G$ is measure preserving, where $\mu(G^t(A))=\mu(A)$, for all Borel sets $A$. Furthermore there is a program that when given an encoding of a fast Cauchy sequence of $t\in\R$ and $x\in\mathcal{X}$, outputs an encoding of a fast Cauchy sequence of $G^tx$.
\end{dff}

\begin{thr}[\cite{Vereshchagin21,Levin74}]
\label{thr:probcons}
Let $P_\rho$, be a family of probability distributions over $\IS$, indexed by $\rho\in\IS$. Assume that there is a Turing machine $T$ such that for all $\rho\in\IS$ computes $P_\rho$ having oracle access to $\rho$. By  ``compute'' we mean all the measures of the cylinder sets  $P_\rho(x\IS)$, can be computed, uniformly in $x\in\FS$. Then there is a constant $c_T>0$ solely dependent on $T$ such that
$$
P_\rho \{ \gamma : \I(\langle\gamma,\rho\rangle:\ch) > m\} < 2^{\I(\rho:\ch)-m+c_T}.
$$
\end{thr}

\begin{lmm}
\label{lmm:measmorphism}
Let $Q:D\subset \mathcal{X}\rightarrow \mathcal{Y}$ be a morphism of computable measure spaces $(\mathcal{X},\mu)$ and $ (\mathcal{Y},\nu)$, with universal tests $\t_\mu$ and $\t_\nu$. There is a $c\in \N$ with the following properties. If $x\in X$ and $\t_\mu(x)<\infty$, then $Q(x)$ is defined and $\t_\nu(Q(x))<\t_\mu(x)+c$.
\end{lmm}
The proof for this lemma is identical to that of the proof of Lemma \ref{lmm:morphism}.
\begin{dff}
\label{dff:defrandmeas}
We update Definition \ref{dff:defrand} to arbitrary measures. For computable measure $\mu$, $\langle \mu\rangle\in\FS$ is the smallest program that can compute $\mu(x\IS)$, uniformly computable $x\in\FS$.  The deficiency of randomness of an infinite sequence $\alpha\in\IS$ with respect to a computable measure $\mu$ over $\IS$ is defined to be $$\D(\alpha|\mu)=\log\sup_n\m(\alpha[0..n]|\langle \mu\rangle)/\mu(\alpha[0..n]).$$ 
By \cite{Gacs21}, $2^\D$ is a lower-computable $\mu$-test, in that $\int_{\IS} 2^{\D(\alpha|\mu)}d\mu(\alpha)=O(1)$, lower computed by program of size $\K(\langle \mu\rangle)$. Thus since $\t_\mu$ is a universal uniform test, $\t_\mu(\alpha)\gem \m(\langle \mu\rangle)2^{\D(\alpha|\mu)}$.
\end{dff}

\begin{thr}[Oscillation of Thermodynamic Entropy]
Let $L$ be the Lebesgue measure over $\R$, $(\mathcal{X},\mu)$ be a computable measure space with computable $\mu(X)$, $\alpha\in\mathcal{X}$, with finite $\I(\alpha:\ch)$. For transformation group $G^t$ acting on $\mathcal{X}$, there is a constant $c$ with $L\{t\in [0,1]:\H_\mu(G^t\alpha) < \log \mu(X)-n\}> 2^{-n-\K(n)-c}$. 
\end{thr}

\begin{prf}
We first assume not. There exists $(G^t,\mathcal{X})$ and computable measure space $(\mathcal{X},\mu)$ and there exists $\alpha\in X$ with finite $\I(\alpha:\ch)$ such that for all $c\in\N$, there exists $n$,  where  
\begin{align*}
L(\{ t\in[0,1] : \H_\mu(G^t\alpha)<\log \mu(\mathcal{X})-n\}) &< 2^{-n-\K(n)-c}\\
L(\{ t\in[0,1] : n-\log \mu(\mathcal{X})<\log\t_\mu(G^t\alpha)\}) &< 2^{-n-\K(n)-c}.
\end{align*}
We sample $2^{n+\K(n)+c-1}$ elements $F$ by choosing a time $t$ uniformly between $[0,1]$. The probability that all samples $\beta\in F$ have $\t_\mu(G^\beta\alpha)\leq n-\log\mu(\mathcal{X})$ is
\begin{align*}
&\prod_{i=1}^{|F|}L\{t\in [0,1]:\log\t_\mu(G^t\alpha)\leq n-\log\mu(\mathcal{X})\}\\
\geq &(1-|F|2^{-n-\K(n)-c})\\
\geq &(1-2^{n+\K(n)+c-1}2^{-n-\K(n)-c})\\
\geq &1/2.
\end{align*}
Let $(\IS,\Gamma)$ be the Cantor space with the uniform measure. The binary representation (see Definition \ref{dff:binary}) creates an isomorphism $(\phi,\phi^{-1})$ of computable probability spaces between the spaces $(\IS,\Gamma)$ and $([0,1],L)$. It is the canonical function $\phi(\gamma)=0.\gamma$. Thus for all Borel sets $A\subseteq [0,1]$, $\Gamma(\phi^{-1}(A))=L(A)$. Since 
$\{t\in [0,1]:\log\t_\mu(G^t\alpha)\leq n-\log\mu(\mathcal{X})\}$ is closed,
\begin{align*}
&L\{t\in [0,1]:\log\t_\mu(G^t\alpha)\leq n-\log\mu(\mathcal{X})\}
=\Gamma\{\gamma\in \IS:\log\t_\mu(G^{\phi(\gamma)}\alpha)\leq n-\log\mu(\mathcal{X})\}.
\end{align*}
So
\begin{align*}
1/2\leq &\prod_{i=1}^{|F|}\Gamma\{\gamma\in\IS:\log\t_\mu(G^{\phi(\gamma)}\alpha)\leq n-\log\mu(\mathcal{X})\}.
\end{align*}
Let $(\delta,\mu_\delta)$ be a binary representation (see Definition \ref{dff:measbinary}), for the computable measure space $(\mathcal{X},\mu)$. Thus $\mu_\delta$ is a computable (not necessarily probability) measure over $\IS$. By Lemma \ref{lmm:measmorphism}, there is a $c'>0$, where
\begin{align*}
&\prod_{i=1}^{|F|}\Gamma\{\gamma\in\IS:\log\t_{\mu_\delta}(\delta^{-1}(G^{\phi(\gamma)}\alpha))\leq n-\log\mu(\mathcal{X})+c'\}\geq 1/2.
\end{align*}
Let $f:\IS\times\IS\rightarrow\IS$, where $f(\gamma,\langle \overrightarrow{\zeta}\rangle) = \delta^{-1}(G^{\phi(\gamma)}\zeta)$. Note, $f(\gamma,\langle \overrightarrow{\zeta}\rangle)$ can be undefined when $\t_\mu(G^{\phi(\gamma)}\zeta)=\infty$, because the morphism $\delta^{-1}$ is only proven to be defined on a constructive $G_\delta$ set of full measure which includes random points. Let $\xi = \langle \overrightarrow{\alpha}\rangle$ be an encoding of a fast Cauchy sequence $ \overrightarrow{\alpha}$ such that $\I(\xi:\ch)<\infty$. The sequence $\xi$ is guaranteed to exist because the assumption of the theorem statement. So
\begin{align*}
&\prod_{i=1}^{|F|}\Gamma\{\gamma\in\IS:\log\t_{\mu_\delta}(f(\gamma,\xi))\leq n-\log\mu(\mathcal{X})+c'\}\geq 1/2.
\end{align*}
By Definition \ref{dff:defrandmeas}, (and  also updating $c'$)
\begin{align*}
&\prod_{i=1}^{|F|}\Gamma\{\gamma\in\IS:\D(f(\gamma,\xi)|\mu_\delta)\leq n-\log\mu(\mathcal{X})+c'+\K(\mu_\delta)\}\geq 1/2.
\end{align*}
Let $\overline{\mu}_\delta(\alpha) = \mu_\delta(\alpha)/\mu_\delta(\IS)$, which is a computable probability measure over $\IS$. 
\begin{align*}
&\prod_{i=1}^{|F|}\Gamma\{\gamma\in\IS:\D(f(\gamma,\xi)|\overline{\mu}_\delta)\leq n+c'+\K(\mu_\delta)\}\geq 1/2.
\end{align*}

Let $\Gamma^{n+c}$ be a computable distribution over the product of $1+2^{n+\K(n)+c-1}$ independent probability measures over $\IS$, encoding into a $\IS$ in the standard way. The first probability distribution gives measure 1 to $\xi$ and the last $2^{n+\K(n)+c}$ probability measures are the uniform distribution $\Gamma$ over $\IS$. So
\begin{align*}
\Gamma^{n+c}(&\textrm{Encoding of $1+2^{n+\K(n)+c-1}$ elements with the first encoded sequence being $\xi$}\\
&\textrm{and the rest of encoded sequences $\beta$ has $\D(f(\beta,\xi)|\overline{\mu}_\delta) \leq n+c'+\K(\mu_\delta)$})\geq 1/2.
\end{align*}
	Let $n^*=\langle n,\K(n)\rangle$. There is an infinite sequence $\eta=\langle n,\K(n),c\rangle\xi$ and a Turing machine $T$, such that $T$ computes $\Gamma^{n+c}$ when given oracle access to $\eta$. By Theorem \ref{thr:probcons}, with the universal Turing machine relativized to $n^*$, and folding the constants together,
	\begin{align*}
	&\Gamma^{n+c}(\{\gamma:\I(\gamma:\ch|n^*)>m\})\\
	<& \Gamma^{n+c}(\{\gamma:\I(\langle \gamma,\eta\rangle:\ch|n^*)\gea m\})\\
	\lem &2^{-m+\I(\eta:\ch|n^*)+c_T}\\
	\lem &2^{-m+\K(n,\K(n),c|n^*)+\I(\xi:\ch|n^*)+c_T}\\
	\lem &2^{-m+\K(c)}.
	\end{align*} 
	Therefore, 
	\begin{align*} \Gamma^{n,c}(\{\gamma:\I(\gamma:\ch|n^*)\gea\K(c)\})&\leq 1/4.
	\end{align*}
	Thus, by probabilistic arguments, there exists $\kappa\in\IS$, such that $\kappa=\langle D,\xi\rangle$, where $D\subset\IS$ and $|D|=2^{n+\K(n)+c-1}$ and each $\beta\in D$ has $\D(f(\beta,\xi)|\overline{\mu}_\delta)\leq n+c'+\K(\mu_\delta)$ and $\I(\kappa:\ch|n^*)\lea \K(c)$. Thus since $\K(f(D,\xi)|\kappa,n^*)=O(1)$ we have $\I(f(D,\xi):\ch|n^*)\lea \I(\kappa:\ch|n^*)\lea \K(c)$. By Theorem \ref{thr:InfExt}, relativized to $n^*$, on the set $D'=f(D,\xi)$ and probability $\overline{\mu}_\delta$, there exists constants $d,f\in\N$ where
	\begin{align}
\nonumber
	m=\log |D|&< \max_{\beta\in D'}\D(\beta|\overline{\mu}_\delta,n^*)+2\I(D':\ch|n^*)+d\K(m|v)+f\K(\overline{\mu}_\delta|n^*)\\
	\nonumber
	m&< \max_{\beta\in D'}\D(\beta|\overline{\mu}_\delta)+\K(n)+2\I(D':\ch|n^*)+d\K(m|n^*)+f\K(\mu_\delta|n^*)\\
	\nonumber
	 &\lea \max_{\beta\in D'}\D(\beta|\overline{\mu}_\delta)+\K(n)+2\K(c)+d\K(m|v)+f\K(\mu_\delta|n^*)\\
	 	\label{eq:h}
	&\lea n+\K(n)+d\K(m|v)+2\K(c)+(f+1)\K(\mu_\delta).		
	\end{align}
	Therefore:
	\begin{align}
	\nonumber
	m&= n+\K(n)+ c-1\\
	\label{eq:mcthermo}
	\K(m|n^*)&\lea \K(c).
	\end{align}  
	Plugging Equation \ref{eq:mcthermo} back into Equation \ref{eq:h} results in
	\begin{align*}
	n+\K(n)+c &\lea n+\K(n)+2\K(c)+d(\K(c)+O(1))+(f+1)\K(\mu_\delta)\\
	c&\lea (2+d)\K(c)+dO(1)+(f+1)\K(\mu_\delta).
	\end{align*}
	
	This result is a contradiction for sufficiently large $c$ solely dependent $\mathcal{X}$, $G$, $\mu$, and the universal Turing machine.\qed
\end{prf}\\

\begin{cor}[Outliers in Dynamics]
Let $L$ be the Lebesgue measure over $\R$, $(\mathcal{X},\mu)$ be a computable probability space, $\alpha\in\mathcal{X}$, with finite $\I(\alpha:\ch)$. For transformation group $G^t$ acting on $\mathcal{X}$, there is a constant $c$ with $L\{t\in [0,1]:\t_\mu(G^t\alpha) > 2^n\}> 2^{-n-\K(n)-c}$. 
\end{cor}

Let $(\mathcal{X},\mu)$ be a  computable probability space. Let $([0,1],L)$ be the computable probability space, where $L$ is the Lebesgue measure. $(\mathcal{X}\times [0,1],\mu\times L)$ is a product computable probability space, which we will be using in the following lemma. It provides a different proof to the same lemma in \cite{Gacs94}.

\begin{lmm}
Let $L$ be the Lebesque measure over $\R$, $(\mathcal{X},\mu)$ be a computable measure space, and $\alpha\in \mathcal{X}$. For transformation group $G^t$ acting on $\mathcal{X}$, there is a constant $c$ where $L\{ t \in [0,1] : \H_\mu(G^t\alpha)< \H_\mu(\alpha)-m \}< 2^{-m+c}$.
\end{lmm}
\begin{prf}
Since
\begin{align*}
 \int_\mathcal{X}\int_{[0,1]}2^{-\H_{\mu\times L}(\alpha,t)}dL(t)d\mu(\alpha)&=\int_\mathcal{X}\int_{[0,1]}\t_{\mu\times L}(\alpha,t)dL(t)d\mu(\alpha)\leq 1,
 \end{align*}
the function $f(\alpha) = \int_{[0,1]}2^{-\H_{\mu\times L}(\alpha,t)}dL(t)$ is a $\mu$-test. So
\begin{align*}
\int_{[0,1]} 2^{-\H_{\mu\times L}(\alpha,t)}dt &=f(\alpha) \lem \t_\mu(\alpha) \eqm 2^{-\H_\mu(\alpha)}.
\end{align*}
So
\begin{align*}
\{ t \in [0,1] : 2^{-\H_{\mu\times L}(\alpha,t)}>2^{m-\H_\mu(\alpha)} \}&\lem 2^{-m}\\
\{ t \in [0,1] : \H_{\mu\times L}(\alpha,t)< \H_\mu(\alpha)-m \}&\lem 2^{-m}
\end{align*}
$\H_{\mu\times L}(\alpha,t)\lea \H_{\mu}(G^t\alpha)$ because 
\begin{align*}
&\int_{[0,1]}\int_\mathcal{X} \t_{\mu}(G^t\alpha)d\mu(\alpha)dL(t)\\
=&\int_{[0,1]}\int_\mathcal{X} \t_{\mu}(\alpha)d\mu(G^{-t}\alpha)dL(t)\\
=&\int_{[0,1]}\int_\mathcal{X}\t_{\mu}(\alpha)d\mu(\alpha)dL(t),\\
=&\int_{[0,1]}1dL(t)\\
\leq&1,
\end{align*}
which means $\t_{\mu}(G^t\alpha)\lem\t_{\mu\times L}(\alpha,t)$ and thus $2^{-\H_{\mu}(G^t\alpha)}\lem 2^{-\H_{\mu\times L}(\alpha,t)}$. Thus 
$$\{ t \in [0,1] : \H_{\mu}(G^t\alpha)< \H_\mu(\alpha)-m \}\lem 2^{-m}.$$
\end{prf}

\begin{cor}
Let $L$ be the Lebesgue measure over $\R$, $(\mathcal{X},\mu)$ be a computable measure space, and $\alpha\in\mathcal{X}$, with finite $\I(\alpha:\ch)$. For transformation group $G^t$ acting on $\mathcal{X}$, there are constants $c_1,c_2>0$ with $2^{-n-\K(n)-c_1}<L\{t\in [0,1]:\H_\mu(G^t\alpha) < \H_\mu(\alpha)-n\}< 2^{-n+c_2}$. 
\end{cor}

\section{Discrete Dynamics}
Theorem \ref{thr:main1} has a discrete version.  Discrete dynamics is modeled by a transform group $G^t$ from Definition \ref{dff:transformgroup}, but with $t\in\mathbb{Z}$, being an integer. We assume there no $\alpha\in\mathcal{X}$ with a finite orbit. Discrete dynamics will visit states with ever increasing $\t_\mu$ score. Given a finite set $D\subset \mathcal{X}$, with $D=\{\alpha_i\}_{i=1}^n$, its mutual information with the halting sequence 
is defined by $\I(D:\ch)=\inf_{\overrightarrow{\alpha_1},\dots,\overrightarrow{\alpha_n}}\I(\langle \overrightarrow{\alpha_1},\dots,\overrightarrow{\alpha_n}\rangle:\ch)$, which is the infimum over all encoded fast Cauchy sequences to members of $D$. Thus $\I(\Omega(\alpha,n):\ch)\lea \I(\alpha:\ch)+\K(n,f)$, using Definition \ref{dff:pointmuthalt}. 
\begin{lmm}
\label{lmm:exoticset}
Given computable measure space $(\mathcal{X},\mu)$, there is a constant $c_{\mathcal{X},\mu}$, with universal uniform test $\t_\mu$, for a finite set $Z\subset \mathcal{X}$ with $n=\ceil{\log|Z|}$, $n< \log\max_{\alpha\in Z}\t_\mu(\alpha) + \I(\langle Z\rangle:\ch) + O(\log \I(\langle Z\rangle:\ch) +\K(n)+c_{X,\mu})$.
\end{lmm}
\begin{prf}
Let $(\IS,\mu_\delta)$ be a binary represention that is isomorphic to computable measure space $(\mathcal{X},\mu)$, with $\delta:(\IS,\mu_\delta)\rightarrow(\mathcal{X},\mu)$. If $\max_{\alpha\in Z}\t_\mu(\alpha)=\infty$, then the lemma is proven. Thus $\delta^{-1}(\alpha)$ is defined for all $\alpha\in Z$. 
 Let $W=\delta^{-1}(Z)\subset\IS$. By Theorem \ref{thr:InfExt} applied to $W$ and $\mu_\delta$, with $s=n-O(1)<\log\sum_{\alpha\in W}2^{\D(\alpha|\mu_\delta)}$, gives 
\begin{align*}
s<&\max_{\alpha\in W}\D(\alpha|\mu_\delta)\,{+}\,\I( W:\ch)+O(\log\I(W:\ch)+\K(s)+c_{\mathcal{X},\mu}).
\end{align*}
Due to Definition \ref{dff:defrandmeas},
\begin{align*}
n<&\max_{\alpha\in W}\log\t_{\mu_\delta}(\alpha)\,{+}\,\I( W:\ch)+O(\log\I(W:\ch)+\K(n)+c_{\mathcal{X},\mu}).
\end{align*}
Since $(\{0,1\}^\infty,\mu_\delta)$ is isomorphic to $(\mathcal{X},\mu)$,
\begin{align*}
n<&\max_{\alpha\in Z}\log\t_\mu(\alpha)\,{+}\,\I( W:\ch)+O(\log\I(W:\ch)+\K(n)+c_{\mathcal{X},\mu}).
\end{align*}
Given any encoding of the fast Cauchy sequences of the members of $Z$, one can compute $W$ with $\delta^{-1}$, thus $\K(W|Z)=O(1)$, so
\begin{align*}
n<&\max_{\alpha\in Z}\log\t_\mu(\alpha)\,{+}\,\I(Z:\ch)+O(\log\I(Z:\ch)+\K(n)+c_{\mathcal{X},\mu}).
\end{align*}
\qed
\end{prf}

\begin{thr}
Let $(\mathcal{X},\mu)$ be a computable measure space and $\alpha\in\mathcal{X}$, with finite $\I(\alpha:\ch)$. For discrete time dynamics $G^t$, there is a $c$ such that $\max_{\gamma\in G^{\{1,\dots,2^n\}}\alpha}\t_\mu(\gamma)>2^{n-O(\K(n))-c}$.
\end{thr}
\begin{prf}
Let $Z_n = G^{\{1,\dots,2^n\}}\alpha$. Lemma \ref{lmm:exoticset}, applied to $(\mathcal{X},\mu)$ and $Z_n$, results in $\gamma\in Z_n$ such that 
\begin{align*}
n <& \log \t_\mu(\gamma) + \I(Z_n:\ch) + O(\log\I(Z_n:\ch)+\K(n)+c_{\mathcal{X},\mu,\alpha}).
\end{align*}
Since $\I(Z_n:\ch)\lea \I(\alpha:\ch)+\K(n)$,
\begin{align*}
n <& \log \t_\mu(\gamma) + \I(\alpha:\ch) + O(\log\I(\alpha:\ch)+\K(n)+c_{\mathcal{X},\mu,\alpha,G}).
\end{align*}
The theorem is proven by noting $\I(\alpha:\ch)<\infty$.
\qed
\end{prf}
\section{Coarse Grained Entropy}

Coarse grained entropy was introduced in \cite{Gacs94} as an update to Boltzmann entropy. The goal was a parameter independent formulation of entropy. It was defined using cells. In this section we define coarse grained entropy with respect to open sets, leveraging the results of \cite{HoyrupRo09}. Let $\Pi(\cdot)$ be a
set of disjoint uniformly enumerable open sets in the computable metric space $\mathcal{X}$.
\begin{dff}[Algorithmic Coarse Grained Entropy] 
$\H_\mu(\Pi_i) = \K(i|\mu)+\log \mu(\Pi_i)$.
\end{dff}
Coarse grained entropy is an excellent approximation of fine grained entropy, as shown by the following two results.
\begin{prp}
Let $(\mathcal{X},\mu)$ be a computable measure space. If $\mu(\Pi_i)$ is uniformly computable and $\alpha\in\Pi_i$ then $\H_\mu(\alpha)\lea \H_\mu(\Pi_i)+\K(\Pi)$.
\end{prp}
\begin{prf}
Let $t(\alpha)= [\alpha\in\Pi_i]\m(\Pi_i)/\mu(\Pi_i)$. $t\in\mathcal{F}$ is lower semi-computable and $\int_{\mathcal{X}}t(\alpha)d\mu(\alpha) = \sum_i \int_{\Pi_i} (\m(\Pi_i)/\mu(\Pi_i)d\mu(\alpha)=\sum_i\m(\Pi_i)\leq 1$. Thus $\t_\mu(\alpha)\gem t(\alpha)$.
\end{prf}$ $\\

\begin{lmm}
\label{lmm:intdom}
For computable measure space $(\mathcal{X},\mu)$, for lower computable function $f\in\mathcal{F}$, and enumerable open set $U$, $\int_Ufd\mu$ is lower computable.
\end{lmm}
\begin{prf}
For a finite union of balls $V=\bigcup_{j=1}^nB_{i_j}$ and an enumerable open set $W=\bigcup_{j=1}^\infty B_{k_j}$ and a computable measure $\mu$, the term $\mu(V\cap W)$ is lower computable. Due to Corollary \ref{cor:compmeas}, the term $\mu(\bigcup\{B:\exists_{s,t}\textrm{ such that }B\subseteq B_{i_s}\textrm{ and } B\subseteq B_{k_t}\})=\mu(V\cap W)$ is lower computable.

The integral of a finite supremum of step functions over $U$ is lower computable by induction. For the base case $\int_Uf_{i,j}d\mu=q_j\mu(B_i\cap U)$ is lower computable by the above reasoning. For the inductive step
$$
\int_U\sup\{f_{i_1,j_1},\dots f_{i_k,j_k}\}d\mu = q_{j_m}\mu\left((B_{i_1}\cup\dots\cup B_{i_k})\cap U\right) +
\int_U\sup\{f_{i_1,j'_1},\dots f_{i_k,j'_k}\}d\mu,
$$
where $q_{j_m}$ is minimal among $\{q_{j_1},\dots,q_{j_k}\}$ and $q_{j'_1}=q_{j_1}-q_{i_m},\dots, q_{j'_k}=q_{j_k}-q_{i_k}$. The first term on the right is lower-computable and by the induction assumption, the last term on the right is lower-computable.
\qed
\end{prf}$ $\\

The following lemma is an update to the Stability Theorem 5 in \cite{Gacs94}, using open sets instead of cells. 
\begin{lmm}
For computable measure space $(\mathcal{X},\mu)$, $\mu\{\alpha\in \Pi_i: \H_\mu(\alpha) < \H_\mu(\Pi_i) - \K(\Pi)-m\}\lem 2^{-m}\mu(\Pi_i)$.
\end{lmm}
\begin{prf}
Let $f(i)=\int_{\Pi_i}\t_\mu(\alpha)d\mu(\alpha)$. By Lemma \ref{lmm:intdom}, the function $f(i)$ is lower computable, and $\sum_if(i)\leq 1$. Thus $f(i)\lem \m(i)/\m(\Pi)$. So
$$
\mu(\Pi_i)^{-1}\int_{\Pi_i}2^{-\H_\mu(\alpha)}d\mu(\alpha) \lem 2^{-\H_\mu(\Pi_i)+\K(\Pi)}.
$$
By Markov inequality,
$$\mu\{\alpha\in \Pi_i: \H_\mu(\alpha) < \H_\mu(\Pi_i) - \K(\Pi)-m\}\lem 2^{-m}\mu(\Pi_i).$$
\end{prf}
\begin{cor}
For computable measure space $(\mathcal{X},\mu)$, $\mu\{\alpha: \H_\mu(\alpha) < \log\mu(\mathcal{X})-m\}\lem 2^{-m}\mu(\mathcal{X})$.
\end{cor}

\end{document}